\documentclass[twocolumn]{aastex62}
\usepackage{natbib}
\usepackage{float}
\usepackage{amssymb}
\usepackage{amsmath}
\usepackage{comment}
\usepackage{ulem}

\newcommand{\ffm}{$f_\textrm{m}$}

\newcommand{\nickel}{$^{56}\textrm{Ni}$}
\newcommand{\wave}{$\sim$}

\newcommand{\angstrom}{{\normalfont\AA}}

\begin{document}
\title{The effect of circumstellar matter on the double-peaked type Ic supernovae and
implications for LSQ14efd, iPTF15dtg and SN 2020bvc}
\author{Harim Jin}
\affil{Department of Physics and Astronomy, Seoul National University, 08826, Seoul, South Korea}
\author{Sung-Chul Yoon}
\affiliation{Department of Physics and Astronomy, Seoul National University, 08826, Seoul, South Korea}
\affiliation{Center for Theoretical Physics (CTP), Seoul National University, 08826, Seoul, South Korea}
\author{Sergei Blinnikov}
\affiliation{Kavli Institute for the Physics and Mathematics of the Universe (WPI), The University of Tokyo Institutes for Advanced Study, The University of Tokyo, 5-1-5 Kashiwanoha, Kashiwa, Chiba 277-8583, Japan}
\affiliation{NRC ``Kurchatov Institute"—ITEP, B.Cheremushkinkaya 25, 117218 Moscow, Russia}
\affiliation{All-Russia Research Institute of Automatics (VNIIA), 127055 Moscow, Russia}

\begin{abstract}

Double peaked light curves are observed for some Type Ic supernovae (SNe Ic)
including LSQ14efd, iPTF15dtg and SN 2020bvc.  One possible explanation of the
first peak would be shock-cooling emission from  massive extended material
around the progenitor, which is produced by mass eruption or rapid expansion of
the outermost layers of the progenitor shortly before the supernova explosion.
We investigate the effects of such circumstellar matter (CSM) on the multi-band
optical light curves of SNe Ic using the radiation hydrodynamics code STELLA.  
Two different SNe Ic progenitor masses at the pre-SN stage (3.93~$M_\odot$ and 8.26~$M_\odot$)
are considered in the SN models. 
The adopted parameter space consists of the CSM mass of $M_\mathrm{CSM} =
0.05 - 0.3 M_\odot$, the CSM radius of $R_\mathrm{CSM} = 10^{13} - 10^{15}$~cm
and the explosion energy of $E_\mathrm{burst} = (1.0 - 12.0)\times10^{51}$~erg.
We also investigate the effects of the radioactive nickel distribution on the
overall shape of the light curve and the color evolution. Comparison of our SN
models with the double peaked SNe Ic LSQ14efd, iPTF15dtg and SN 2020bvc
indicate that these three SNe Ic had a similar CSM structure (i.e.,
$M_\mathrm{CSM} \approx 0.1 - 0.2 M_\odot$ and $R_\mathrm{CSM} = 10^{13} -  
10^{14}~\mathrm{cm}$), which might imply a common mechanism for the CSM
formation.  The implied mass loss rate of $\dot{M} \gtrsim  
1.0~M_\odot~\mathrm{yr^{-1}}$ is too high to be explained by the previously
suggested scenarios for pre-SN eruption, which calls for a novel mechanism.   

\end{abstract}

\section{Introduction}

Mass loss from stars can occur through multiple channels like standard
radiation-driven steady winds, pulsation-driven winds, episodic eruptions and
binary interactions.  Mass-loss has a great influence on the evolution of
massive stars and the resultant core-collapse supernova (SN)
types~\citep[e.g.,][]{Smith2014}. While the progenitors of Type II SNe (SNe II)
have a considerable amount of hydrogen in their envelopes, the progenitors of
Type Ib and Ic SNe (SNe Ib/Ic) are supposed to be Wolf-Rayet (WR) stars or
naked helium stars of which the hydrogen envelopes have been stripped off via
mass loss~\citep[e.g.,][]{Yoon2015}. 

If strong mass loss occurred shortly before a SN explosion, it  would create a
thick layer of circumstellar matter (CSM) around the SN progenitor.  The
interaction of such a CSM layer and the SN ejecta would have a significant
impact on the SN light curve and spectra.  The most notable examples are the
interacting supernovae like SNe IIn and SNe Ibn~\citep[e.g.,][]{Blinnikov2017,
Smith2017,Moriya2018a}. Many recent studies on ordinary SNe IIP also report evidence for
the presence of massive CSM around the progenitors~\citep[e.g.,][]{Gonzalez2015,
Khazov2016, Forster2018}. This implies that a large fraction of SN IIP
progenitors would undergo strong enhancement of mass loss at the pre-SN stage,
for which various mechanisms have been proposed in the
literature~\citep[e.g.,][]{Yoon2010, Quataert2012, Woosley2015, Fuller2017}.

Compared to the case of SN IIP progenitors that are mostly red supergiants, SNe
Ib/Ic progenitors are compact and would need more energy for mass
ejection.  However, some recent theoretical studies predict a significant mass
loss enhancement at the pre-SN stage from helium star progenitors by wave
heating~\citep{Fuller2018}, rapid rotation~\citep{Aguilera2018} or silicon
flashes~\citep{Woosley2019}\footnote{The pulsational pair-instability from very
massive helium stars ($34 \lesssim M_\mathrm{He} \lesssim 62~M_\odot$;
\citealt{Woosley2017}) is another possibility for strong mass ejection at the
pre-SN stage. But in the present study, our discussion only focuses on
relatively low-mass helium star progenitors that would undergo core-collapse having a final mass less than about
10 $M_\odot$.}. This would be related to the narrow emission lines of SNe Ibn
and the unusually bright early-time emission of some SNe Ib including SN 2008D
and LSQ13abf.

The presence of massive CSM might also be responsible for the first peak of
several double-peaked superluminous SNe Ic (e.g., LSQ14bdq;
\citealt{Nicholl2015, Nicholl2016}) and  peculiar SNe Ic like SN
2006aj~\citep[e.g.,][]{Modjaz2006}, iPTF15dtg~\citep{Taddia2016, Taddia2019}
and SN 2020bvc~\citep{Ho2020, Rho2020}. The magnetar scenario is often invoked
to explain such an unusual SN Ic.  Given that rapid rotation is a necessary condition
for the production of a magnetar, rotationally-driven rapid mass loss  during
the final evolutionary stage might commonly occur for magnetar
progenitors~\citep{Aguilera2018}.  Alternatively, the double peak feature could
be explained by the magnetar model where the shock driven by  the magnetar
energy breaks out the already expanding SN ejecta~\citep{Kasen2016}.   

On the other hand, no double-peaked light curve has been found for most of the
ordinary SNe Ic that are powered by radioactive $^{56}$Ni. To our knowledge,
the SN Ic LSQ14efd is the only ordinary SN Ic (in terms of energy, ejecta and
nickel mass; see below) that shows a signature of the double peaked light curve
(i.e., bright post-breakout emission; \citealt{Barbarino2017}).  Given that
strong pre-SN mass loss would be a likely reason for this
double peak feature and that the mass loss mechanism might be different from
the case of the magnetar-powered SNe, it would be worth investigating the
effect of CSM on the early-time light curves of SNe Ic to infer the physical
properties of the CSM around the LSQ14efd progenitor and to provide a
theoretical constraint for future observations of SNe Ic.  For this purpose, we
present multi-color SN Ic light curve models calculated with the radiation
hydrodynmics code {\sc STELLA} considering a thick CSM environment around the
progenitor and apply the results to LSQ14efd, of which the
photometric data are given by \citet{Barbarino2017}.

Although our original motivation is to explain the optical light curve of
LSQ14efd, we also apply our results to two other double peaked SNe Ic:
iPTF15dtg and SN 2020bvc.  Intriguingly, we find that these double-peaked SNe
Ic of our sample had similar CSM properties in terms of CSM mass and radius,
which might imply a common mechanism of pre-SN mass ejection. 

In Section 2, we describe the SN Ic progenitor models, the considered parameter
space and the numerical method.  In Section 3, we show the effects of different
parameters of CSM on the light curves and color evolution of SNe Ic. In Section
4, we apply our result to LSQ14efd, iPTF15dtg, and SN 2020bvc, and discuss its
implications for the mass loss mechanism.  In Section 5, we present our
conclusions.

\begin{table}
\begin{center}
\caption{Progenitor model properties}\label{tab1}
\begin{tabular}{ c |  c c c c c c}
\hline
Model   &  $M$             & $R$           & $m_\mathrm{He, env}$ & $Y_\mathrm{s}$   & $M_\mathrm{Fe}$  & $E_\mathrm{bind}$  \\ 
       &  [$M_\odot$]      &  [$R_\odot$]  & [$M_\odot$]          &                  & [$M_\odot$]  &  $10^{51}$~erg \\
\hline
\hline
 4P     &     3.93         & 0.77          &  0.06                &   0.49           & 1.44     & 0.3       \\
 8P     &     8.26         & 0.25          &  0.08                &   0.15           & 1.85     & 1.5       \\
\hline
\end{tabular}
\tablecomments{$M$: total mass of the progenitor model;
$R$: Radius; $m_\mathrm{He, env}$: integrated helium mass for the region above the iron core; $Y_s$: surface helium abundance; 
$M_\mathrm{Fe}$: iron core mass that corresponds to the adopted mass cut; $E_\mathrm{bind}$: binding energy.} 
\end{center}
\end{table}

\section{Modeling} \label{sec:style}

\begin{figure}
\plotone{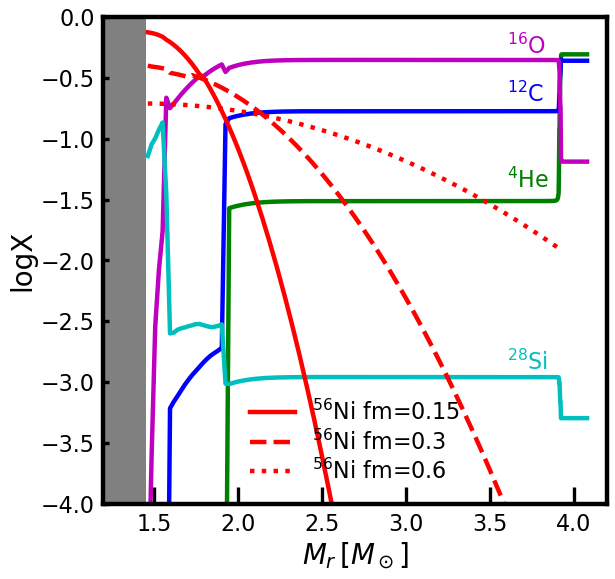}
\plotone{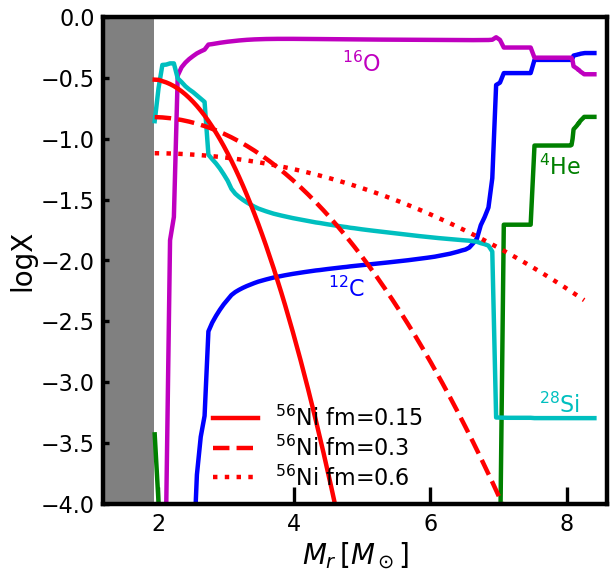}
\plotone{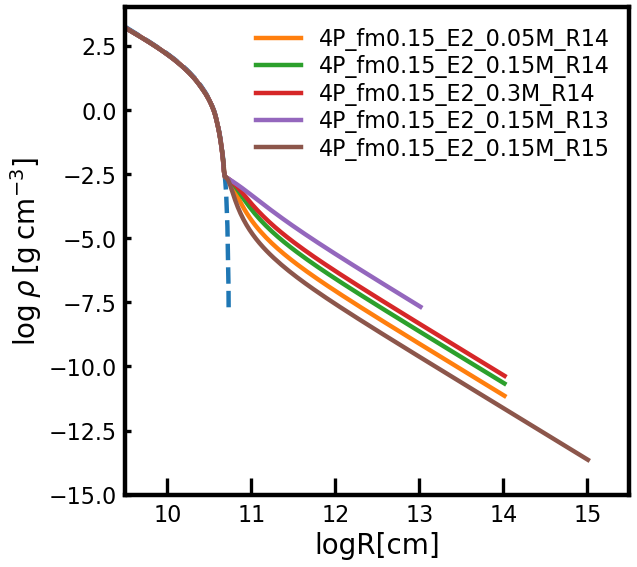}
\caption{Chemical compositions of a 4P model (top) and a 8P model (middle) with  the same nickel distribution parameter, \ffm = 0.15. For comparison, nickel profiles with different \ffm  are also shown on the figure. Density profiles of a representative set of 4P models with different CSM structure (bottom). The dashed line shows the density profile of the original 4P progenitor.}\label{fig:ini}
\end{figure}

Our progenitor models for the SN simulations are helium poor stars with final
masses of 3.93~$M_\odot${}(4P model) and 8.26~$M_\odot${}(8P model) and their properties
are given in Table~\ref{tab1}. These models are obtained by evolving helium stars of
7.0~$M_\odot${} and 15~$M_\odot${} at the initial metallicity $Z_\mathrm{init} = 0.02$,
respectively, with the {\sc MESA} code~\citep{Paxton2011, Paxton2013,
Paxton2015, Paxton2018}. Here we adopt step-overshooting with an overshooting
parameter of 0.1$H_P$ where $H_P$ is the local pressure scale height at the
outer boundary of the helium burning convective core. We use the Wolf-Rayet
mass-loss rate prescription by \citet{Nugis2000} until core helium exhaustion
and a fixed mass-loss rate of $10^{-4}~M_\odot~\mathrm{yr^{-1}}$ during the
later evolutionary stages. The total amounts of helium retained in the outer
region above the iron core are only 0.06~$M_\odot${} and 0.08~$M_\odot${} in 4P and 8P
models respectively, and therefore these models are suitable for SNe Ic rather
than SNe Ib. See also Figure~\ref{fig:ini} for the chemical composition of the
progenitor models.

To calculate the SN models, we use the one-dimensional multi-group radiation
hydrodynamics code {\sc STELLA} \citep{Blinnikov1998, Blinnikov2000,
Blinnikov2006, Blinnikov2011}. It solves a set of time-dependent radiative
transfer equations coupled with the hydrodynamics equations.  The covered
wavelength range  in the calculations is  $5 \times 10^{4} - 10^{-3}$
\angstrom{}, for which 109 wavelength bins are used.  The ionization levels and
excitation levels are obtained with the  assumption of local thermodynamic
equilibrium. Both scattering and absorption are considered in
the opacity treatment~\citep{Blinnikov1998, Kozyreva2020}.  The effects of
fluorescence and non-LTE that are not included in {\sc STELLA} would affect the
light curve especially when the radioactive \nickel{} is present near the
photosphere~\citep[see][for a detailed discussion]{Blinnikov1998}.  
For example, fluorescence would possibly make the SN color bluer than our model prediction
when the light curve is dominated by \nickel{} heating. 
However,
these effects only play a minor role  in the early-time light curve dominated
by the shock cooling emission from the interaction between CSM and SN ejecta,
which is the main concern of this study.  Multi-dimensional effects might also
affect the precise determination of SN parameters. The full description of
fluorescence, non-LTE effects as well as multi-dimensional effects will be
available in a future version of STELLA \citep{Potashov2017,Panov2018}.

The SN explosion is treated as a thermal bomb at the mass cut, which
corresponds to the iron core mass in this study (see Table~\ref{tab1}).  Our
progenitor models are mapped into the {\sc STELLA} code and 250 mass zones
including 80 zones in the CSM are used for SN simulations. Readers are referred
to \citet{Blinnikov2011} and references therein for details of the {\sc STELLA}
code and to \citet{Yoon2019} for a recent example of the use of {\sc STELLA}
for modeling SNe Ib/Ic.

\subsection{Nickel distribution}

We do not calculate the explosive nucleosyntheis, which is not the subject of
this work, and instead put a certain amount of radioactive $^{56}$Ni in the
input progenitor models.  We assume a nickel distribution which follows a
Gaussian profile as in \citet{Yoon2019}: 
\begin{equation}
\label{eq1}
X_\textrm{Ni}(M_r)=A\textrm{exp}\left(-\left[\frac{M_r-M_\textrm{Fe}}{f_\textrm{m}(M_\textrm{tot}-M_\textrm{Fe})}\right]^2\right).
\end{equation}
Here, $A$ is the normalization factor, $M_r$ the mass coordinate,
$M_\textrm{Fe}$ the iron core mass, $M_\textrm{tot}$ the total mass of the
progenitor, and $f_\textrm{m}$  the $^{56}$Ni distribution parameter.  For our
fiducial models, we fix the total \nickel{} mass to  0.25~$M_\odot${}, which is
the inferred value for LSQ14efd~\citep{Barbarino2017}.  This amount of
$^{56}$Ni is within the typical range of the $^{56}$Ni mass distribution of
ordinary SNe Ic~\citep{Anderson2019}.  We calculate SN models using different
values of $f_\mathrm{m}$, which determines the degree of $^{56}$Ni mixing in
the SN ejecta: the $^{56}$Ni distribution becomes flatter for a larger
$f_\mathrm{m}$ and vice versa as shown in Figure \ref*{fig:ini}.

\subsection{CSM structure}

We assume that the CSM has the density profile of  $\rho_{\textrm{CSM}} =
\dot{M} / 4 \pi v_{\textrm{wind}} r^2$, with the standard  $\beta$-law wind
velocity profile:  $v_{\textrm{wind}}(r) = v_0 + (v_\infty - v_0)\left(1-\frac
{R_0} {r} \right)^\beta$. Here $\dot{M}$ is the mass-loss rate of the
progenitor, $v_0$ is the wind velocity at the progenitor surface, $v_\infty$ is
the terminal velocity, $R_0$ is the radius of the progenitor, and $r$ is the
distance from the center of the progenitor. The density profile would follow $r^{-2}$ after the wind is accelerated in a transition layer. According to \citet{Fuller2018},
hydrogen-poor stars are predicted to emit wave-driven outbursts with terminal
velocities of a few 100~$\mathrm{km~s^{-1}}$. We  fix the terminal
velocity to $200~\mathrm{km~s^{-1}}$ in this study. Note that
there exists degeneracy between $\dot{M}$ and $v_\infty$:
a larger $v_\infty$ would imply a larger $\dot{M}$ for a given CSM mass, 
which should be kept in mind when we discuss our result. 

For the wind velocity parameter $\beta$, we assume $\beta = 3.0$.  Although
this value can affect the early-time light curves of Type IIP supernovae
significantly~\citep{Moriya2018b}, our SN Ic models depend on $\beta$ very
weakly because of the small radii of the progenitors. The CSM mass can vary by
a factor of 5 to 500 for $\beta =  1 \cdots 5$ for red supergiant SN progenitors
\citep{Moriya2018b} but only by 3\%  for our models. 

Besides, we assume that the chemical composition of the CSM follows that of the
outermost part of the progenitor. It would be possible that the CSM is more
helium-rich than the progenitor surface. However, our test calculations
indicate that the early-time light curves are not meaningfully affected for our
considered parameter space even if a helium-rich composition is adopted,
although it might be important for the details of early-time spectra.   

In the bottom panel of Figure \ref{fig:ini}, the density profile of the 4P
progenitor is presented with the blue dashed line.  The density decreases very
steeply near the surface of the progenitor. The SN shock is rapidly accelerated
in this region, making the time step very small. To avoid this
numerical difficulty, we take off $4\times10^{-5}M_\odot$ of the outermost
layer of the progenitor when we attach CSM in our models. 
As
discussed below (Section~\ref{sect:rhoprofile}), the choice of the CSM inner
boundary is not important for the conclusions of this study. 

Note also CSM in our models can also be considered as an extended outward
moving envelope created by an energy injection during the final evolutionary
stage.  The difference between an outward moving envelope and wind matter would be just that an envelope
is gravitationally bound to the progenitor while the wind matter is not. As long as the CSM
velocity is much lower than the SN shock velocity, wind matter and an extended envelope
would not lead to a difference in the resulting light curve as long as the
density profile is not very much different (see Section~\ref{sect:rhoprofile}). 

We also calculate some models with a very small amount of CSM  (0.001\% of the
progenitor mass) with $R = 10^{14}~\mathrm{cm}$ for comparison. In this case,
the CSM would correspond to the wind material from an ordinary line-driven
Wolf-Rayet wind having $\dot{M} \sim 10^{-5}~M_\odot~\mathrm{yr^{-1}}$.  For
convenience, these models are denoted by `no-CSM' as the effect of CSM on the
light curve is practically negligible in this case.

\subsection{Considered parameter space}

We focus on the effects of four parameters for a given progenitor mass: CSM
mass ($M_\textrm{CSM}$), CSM radius ($R_\textrm{CSM}$), nickel distribution
($f_\textrm{m}$), and explosion energy ($E_\textrm{burst}$).  In our fiducial
models, we consider $M_\textrm{CSM} = 0.05 \cdots 0.3 M_\odot$, $R_\textrm{CSM}
=  10^{13} \dots 10^{15}~\mathrm{cm}$, $f_\mathrm{m} =$ 0.15, 0.3, and 0.6.  We
consider the explosion energy of $E_\mathrm{burst} = (1 \cdots 3)
\times10^{51}~\mathrm{erg}$.  for 4P models and $E_\mathrm{burst} = (5 \cdots
12) \times10^{51}~\mathrm{erg}$ for 8P models.  The \nickel{} mass is set to
0.25~$M_\odot$ and 0.40~$M_\odot$ for reproducing the main peaks of
LSQ14efd/iPTF15dtg and SN 2020bvc, respectively. 

For simplicity, each model is referred to as
xP\_$f_\textrm{m}$\_$E_\textrm{burst}$\_ $M_\textrm{CSM}$\_log$R_\textrm{CSM}$ in the figures. 
For example, 4P\_fm0.15
\_E2\_0.15M\_R13 denotes the SN model with the 4P
progenitor,  $f_\textrm{m}$=0.15, $E_\textrm{burst}$=2.0B,
$M_\textrm{CSM}$=0.15$M_\odot$, and  log$R_\textrm{CSM}$ [cm] = 13.


\section{Results}

\begin{figure}
    \centering
    \includegraphics[width=0.45\textwidth]{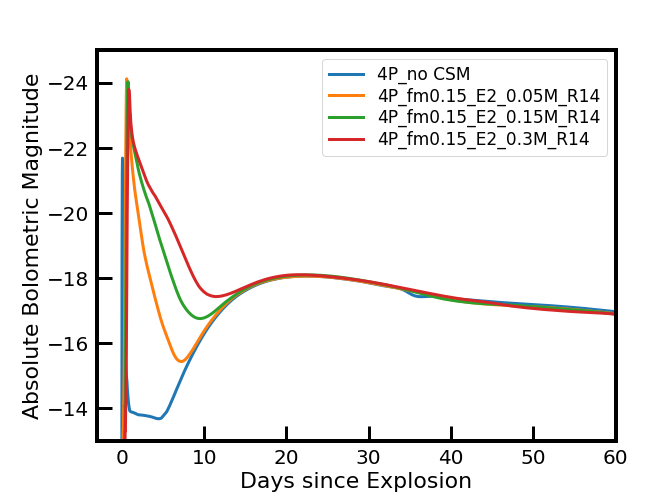}
    \caption{Bolometric light curves of 4P models with $f_\mathrm{m} = 0.15$, $R_\mathrm{CSM} = 10^{14}$~cm, and $E_\mathrm{burst} = 2$B for different CSM 
masses: no-CSM (blue), 0.05~$M_\odot$ (orange), 0.15~$M_\odot$ (green), and 0.30~$M_\odot$ (red brown). 
    \label{fig:bol}}
\end{figure}

\begin{figure*}
    \centering
    \includegraphics[width=1\textwidth]{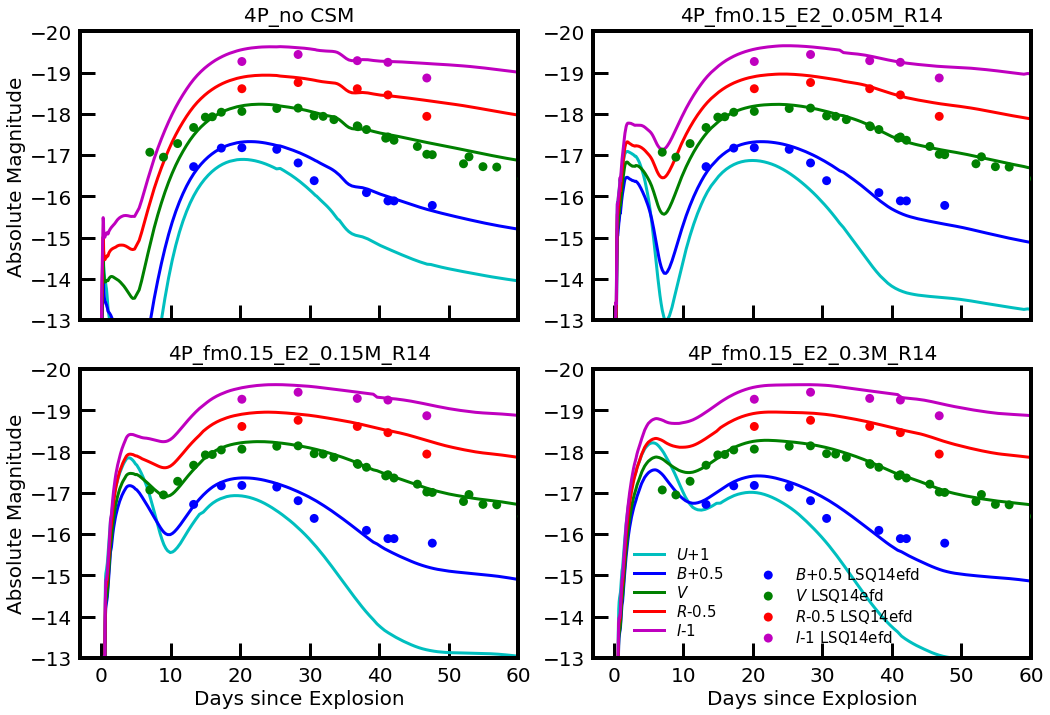}
\caption{Multicolor light curves of models with and without CSM. Each panel shows models with different CSM masses. Different bands are presented using different colors as indicated by the legend in the bottom-right panel. The explosion date is chosen by mathcing
the $V$-band light curve around the main peak with the model.
    \label{fig:ubv}} 
\end{figure*}

\begin{figure}
    \centering
    \includegraphics[width=0.45\textwidth]{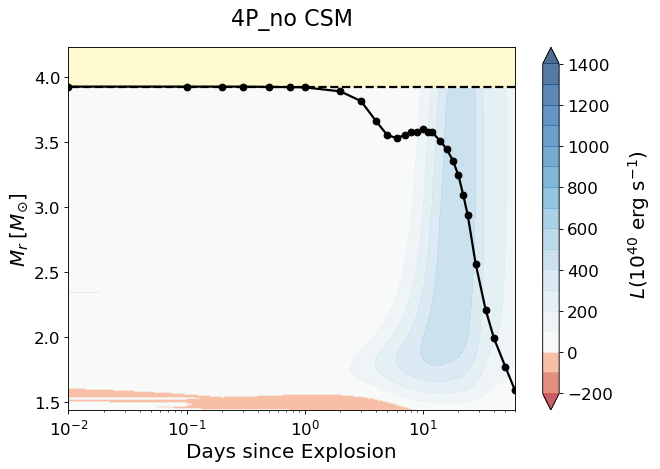}
    \includegraphics[width=0.45\textwidth]{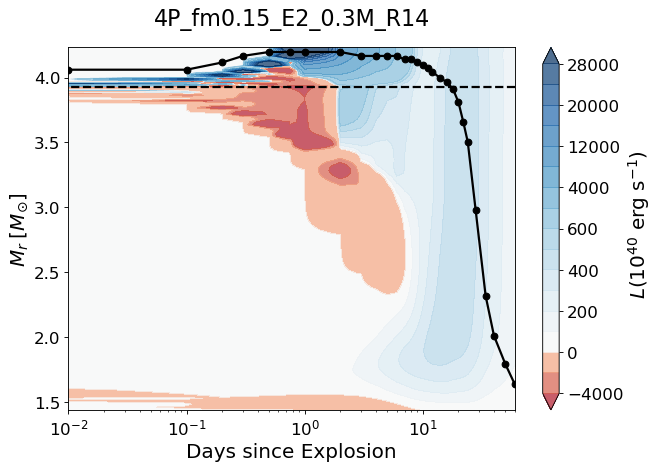}
\caption{Luminosity contour maps of SN models with different CSM masses, 4P\_no-CSM (top) and 4P\_fm0.15\_E2\_0.3M\_R14 (bottom). The points in each epoch denote the photospheres defined as the location where the Rosseland optical depth being 2/3. 
The dashed line represents the surface of the progenitor, which corresponds to the inner boundary of the CSM in the lower panel.
    \label{fig:lum}} 
\end{figure}

\begin{figure}
    \centering
    \includegraphics[width=0.45\textwidth]{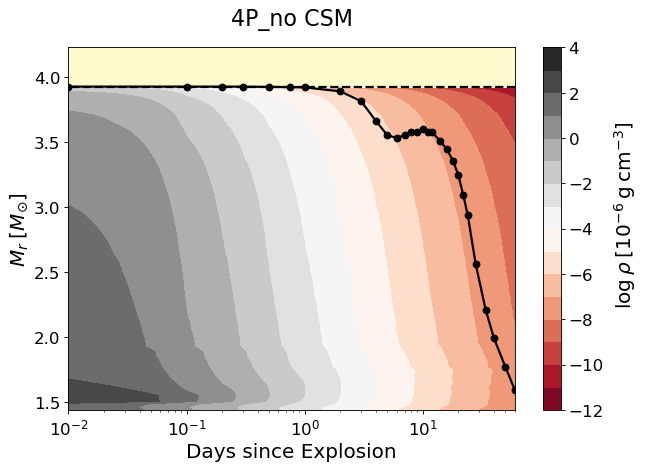}
    \includegraphics[width=0.45\textwidth]{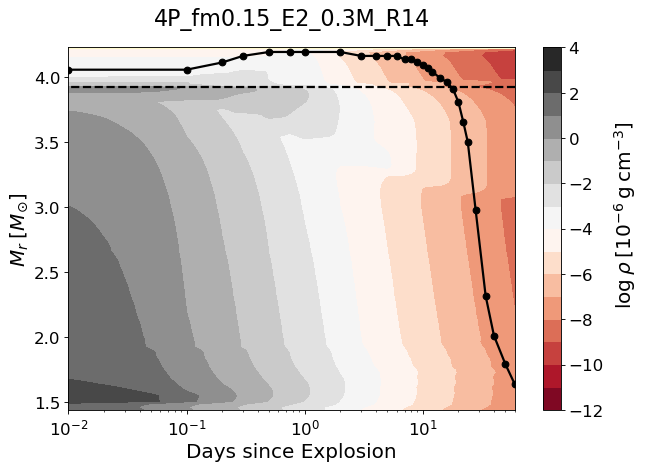}
\caption{Density contour maps of SN models with different CSM masses, 4P\_no-CSM (top) and 4P\_fm0.15\_E2\_0.3M\_R14 (bottom). The points in each epoch denote the photospheres defined as the location where the Rosseland optical depth being 2/3. 
The dashed line represents the surface of the progenitor, which corresponds to the inner boundary of the CSM in the lower panel. 
    \label{fig:den}} 
\end{figure}

\begin{figure}
    \centering
    \includegraphics[width=0.45\textwidth]{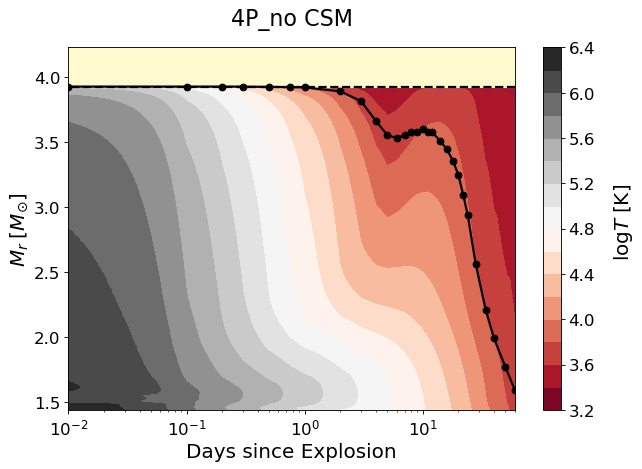}
    \includegraphics[width=0.45\textwidth]{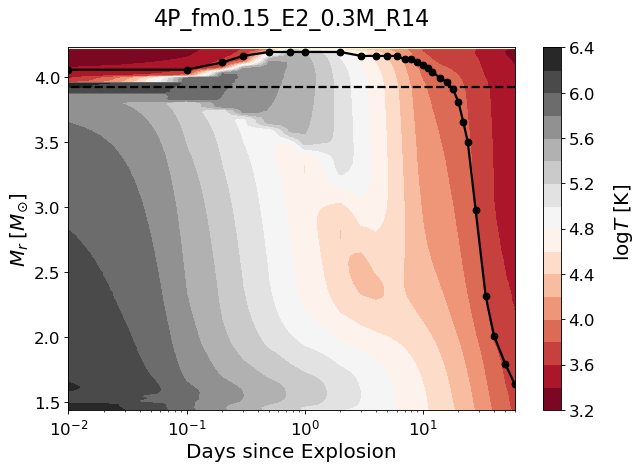}
\caption{Temperature contour maps of SN models with different CSM masses, 4P\_no-CSM (top) and 4P\_fm0.15\_E2\_0.3M\_R14 (bottom). The points in each epoch denote the photospheres defined as the location where the Rosseland optical depth being 2/3. 
The dashed line represents the surface of the progenitor, which corresponds to the inner boundary of the CSM in the lower panel.
    \label{fig:tem}} 
\end{figure}

\begin{figure}
    \centering
    \includegraphics[width=0.45\textwidth]{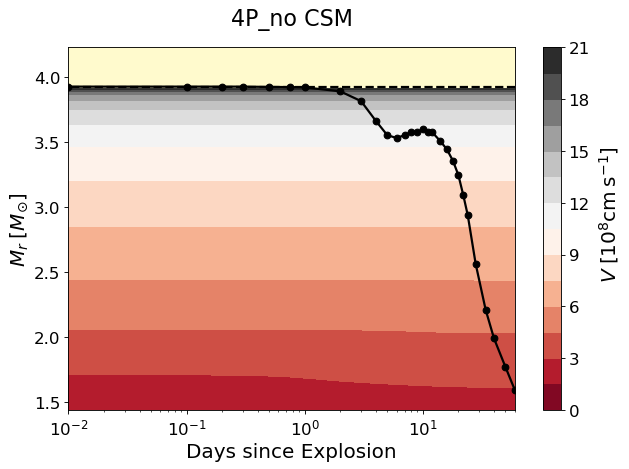}
    \includegraphics[width=0.45\textwidth]{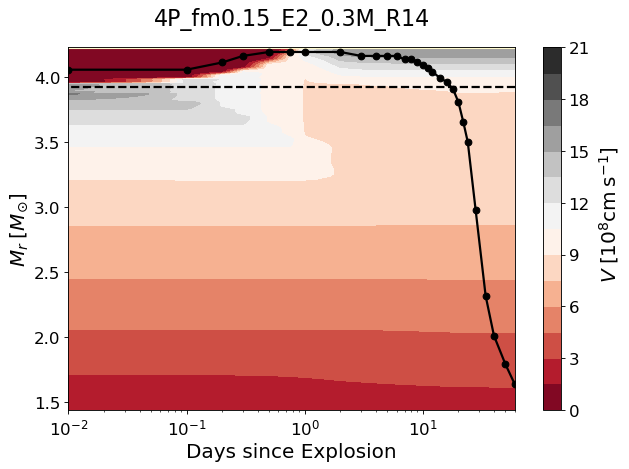}
\caption{Velocity contour maps of SN models with different CSM masses, 4P\_no-CSM (top) and 4P\_fm0.15\_E2\_0.3M\_R14 (bottom). The points in each epoch denote the photospheres defined as the location where the Rosseland optical depth being 2/3. 
The dashed line represents the surface of the progenitor, which corresponds to the inner boundary of the CSM in the lower panel.
    \label{fig:vel}} 
\end{figure}

\begin{figure*}
\includegraphics[width=0.95\textwidth]{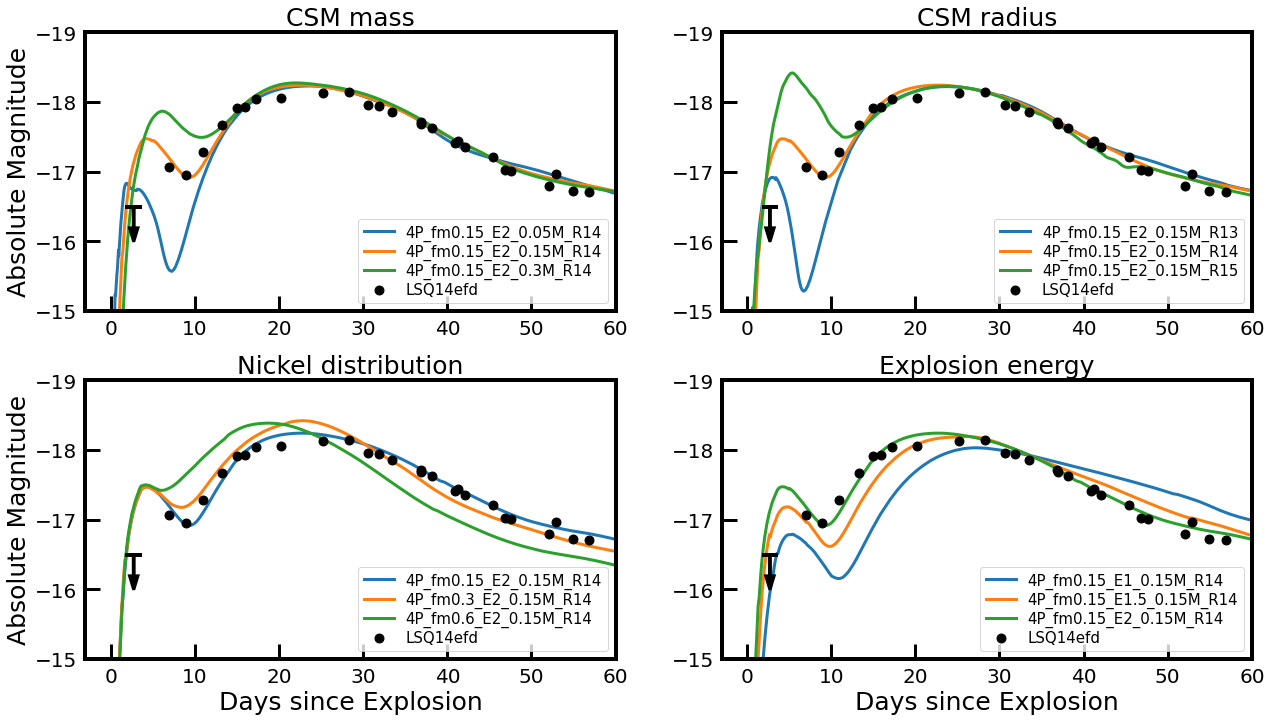}
\caption{$V$-band light curves of 4P models.
Each panel shows models with different CSM masses (upper-left), different
CSM radii (upper-right), different $f_\textrm{m}$ (lower-left), and different
$E_\textrm{burst}$ (lower-right). To see the effects of each parameter, three
models were chosen respectively and drawn in solid lines as indicated by the legends in each panel. Black downward arrows indicate the pre-explosion limit.
The explosion date is chosen by matching the observed $V$-band light curve around the main peak with our fiducial model 4P\_fm0.15\_E2\_0.15M\_R14.
 }\label{fig:lig}
\end{figure*}

\begin{figure*}
        \includegraphics[width=1\textwidth]{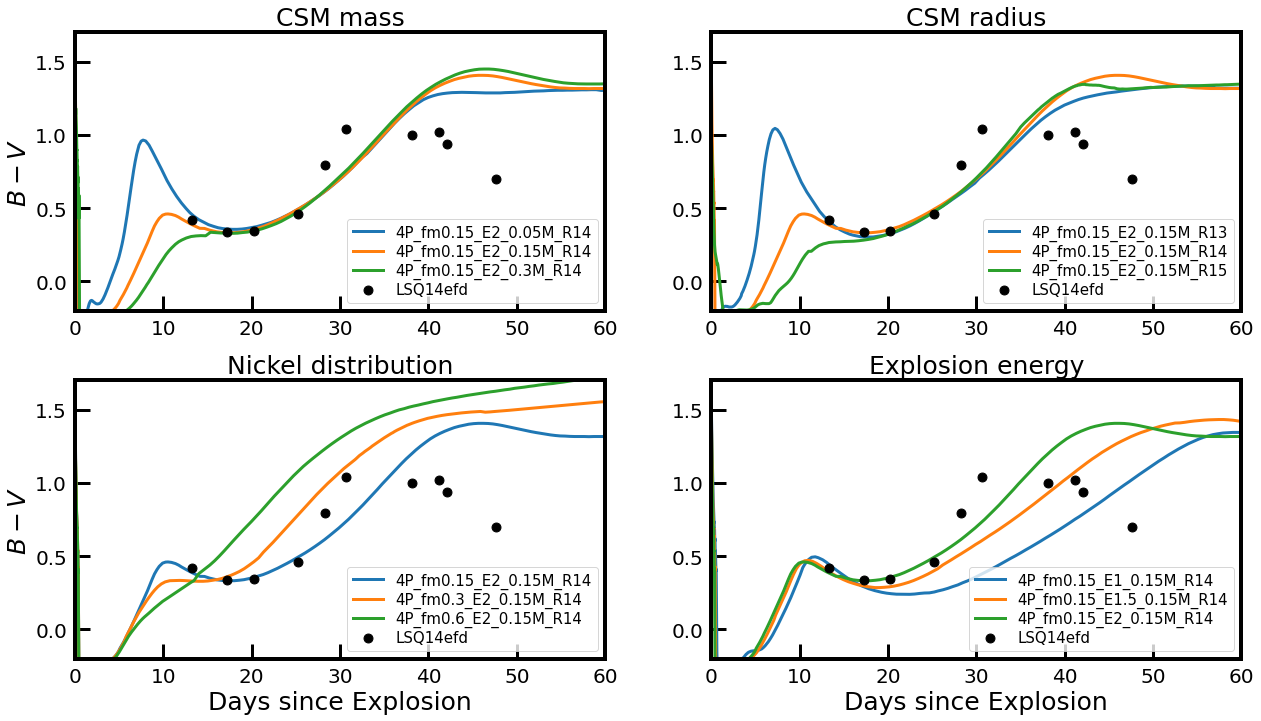}
\caption{Evolution of the $B-V$ color of the 4P SN models presented in Figure~\ref{fig:lig}. Each panel shows the models with different CSM masses (upper-left), different CSM radii (upper-right), different $f_\textrm{m}$ (lower-left), and different $E_\textrm{burst}$ (lower-right). To see the effects of each parameter, three models were chosen respectively and drawn in solid lines as indicated by the legends in each panel. }.
    \label{fig:col}
\end{figure*}

\subsection{General characteristics}\label{sect:Mcsm}

In STELLA, SN shock is initiated by a thermal bomb and begins to
propagate outward from the assumed mass cut. The density profile is steeper
than $r^{-3}$ at the outermost layers of the progenitor (Figure~\ref{fig:ini}) thus
the shock moving forward in these layers is accelerated until it reaches the
surface of the progenitor~\citep{Nadezhin1965,Grasberg1981,Blinnikov2011}. As the forward shock enters CSM, in which the
density profile follows $r^{-2}$, it decelerates and an inward-moving reverse shock is
created. These shocks convert a significant fraction of SN
kinetic energy into internal energy, leading to a bright shock-cooling
emission.  

As an example,  we present bolometric light curves of our  4P models with
$f_\mathrm{m} = 0.15$, $E_\mathrm{burst} = 2.0$B, and $R_\mathrm{CSM} =
10^{14}~\mathrm{cm}$ for various CSM masses in Figure~\ref{fig:bol} and the
corresponding multi-color light curves (i.e., in $U$, $B$, $V$, $R$ and $I$
bands) in Figure~\ref{fig:ubv}.

After the shock breakout, the bolometric
luminosity of the no-CSM model drops very rapidly until the post-breakout
plateau phase is reached~\citep[See][for a detailed discussion on this
short-lived plateau phase of SNe Ib/Ic]{Dessart2011}.  By contrast, for the
models with CSM, it remains brighter and decreases more slowly for several days
(e.g., $\sim$ 10 days for $M_\mathrm{CSM} = 0.15 M_\odot$). The main power
source  during this period is the interaction of the SN shock and the CSM.  We
refer to this period as the `interaction-powered phase (IPP)' following
\citet{Moriya2011}. After the IPP, the light curve is dominated by the energy
due to the radioactive decay of \nickel{}, and we refer to this phase as the
`\nickel{}-powered phase (NPP)'.

To better understand the role of CSM in the IPP,  we compare the evolution of
the local luminosity ($L_r$), density ($\rho_r$), temperature ($T_r$) and
velocity ($v_r$) of the 4P model having $M_\mathrm{CSM} = 0.3 M_\odot$ to those
of the no-CSM 4P model in
Figures~\ref{fig:lum},~\ref{fig:den},~\ref{fig:tem}~and~\ref{fig:vel}.  In the
CSM model, the propagation of the forward and reverse shocks can be traced by
the positive and negative luminosity peaks in the bottom panel of
Figure~\ref{fig:lum}.  No aftereffect of the shocks is seen for the no-CSM
model since there is barely an interplay of the shock and the wind matter. 

The location of the photosphere is also greatly affected by the presence of
CSM. As seen in Figure~\ref{fig:den}, in the CSM model the photosphere
(defined by the Rosseland mean opacity) initially moves upward in the CSM in
the mass coordinate until $t \simeq 1.0$~d.
The photosphere  gradually moves
downward thereafter along with the SN ejecta expansion.

The shocked layers are heated up and the outermost layers in the CSM model
remain much hotter than in the no-CSM model from $t= 0.6$~d when the forward
shock breaks out the CSM (Figure~\ref{fig:tem}).  These shocked layers in the
CSM model have a significantly lower velocity compared to the no-CSM model
(Figure~\ref{fig:vel}).  Note also that a dense shell is formed at $M_r \approx 
3.2 - 3.9 M_\odot$ as the reverse shock sweeps up this region.  

These effects of the shock become more prominent for a larger CSM mass; 
a larger amount of the kinetic energy is transformed into the internal energy
as more layers are shock-heated and decelerated. Furthermore the lower expansion velocity makes the expansion cooling
less efficient and the photon diffusion time longer. 
These factors make the IPP longer and the bolometric and optical luminosities
during the IPP brighter for a larger CSM mass as seen in
Figures~\ref{fig:bol} and~\ref{fig:lig}.  

The CSM also has a great impact on the SN color.  Given that the photosphere
during the IPP is hotter for a larger CSM mass, the color of the SN during the
IPP becomes bluer as seen in Figure~\ref{fig:col} (see the upper left panel).


\citet{Yoon2019} showed that, without CSM, the early-time color evolution of a
SN Ib/Ic sensitively depends on the \nickel{} distribution in the SN ejecta.
Stronger mixing of \nickel{} into the outermost layers would lead to a bluer
color in the earliest days followed by a monotonic reddening during the
photospheric phase, while fairly weak mixing leads to three distinct phases of
initial reddening, blueward evolution, and reddening again. In addition, a
strong \nickel{} mixing tends to suppress the post-breakout emission that would
otherwise appear during early times~\citep{Dessart2012, Piro2013, Yoon2019}. Our CSM
models in Figures~\ref{fig:col} indicate, however, that a monotonic reddening
can also be realised with a sufficient amount of CSM (e.g., for the cases of
4P\_fm0.15\_E2\_0.3M\_R14 and 4P\_fm0.15\_E2\_0.15M\_R15 in Figure~\ref{fig:col}), even
if the \nickel{} mixing is weak (i.e., $f_\mathrm{m} = 0.15$).  
 
On the other hand, the heat due to \nickel{} in the inner region diffuses
outward while the photosphere moves down toward the \nickel{}-heated region as
can be seen in Figure~\ref{fig:lum}.  Once the luminosity is dominated by this
\nickel{} heating, the IPP ends and the NPP begins.  The light curve during the
NPP is determined by the total \nickel{} mass and its distribution and becomes
almost independent of the CSM structure (Figures~\ref{fig:bol}
and~\ref{fig:lig}). 

\subsection{The effects of various parameters}

\begin{figure*}
\includegraphics[width=0.9\textwidth]{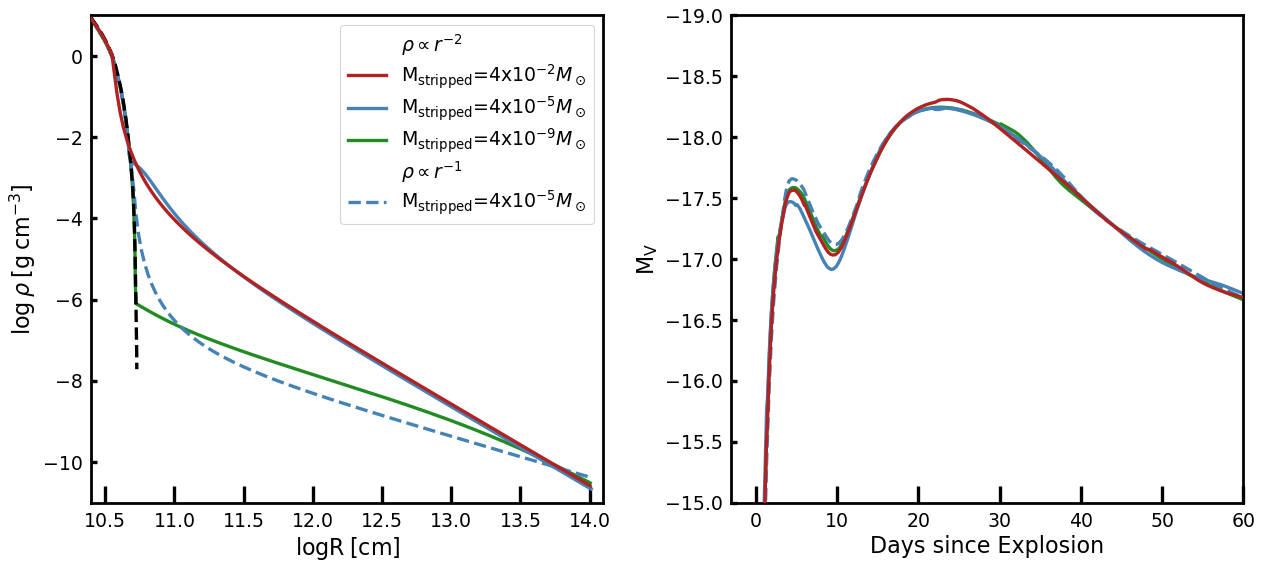}
\caption{Density profiles of models with different $M_\mathrm{stripped}$ (i.e.,
different choices of the boundary between the progenitor star and the CSM; see
the text for details) and density profile laws  (left) and  the corresponding
$V$-band light curves (right). All other parameters are same with our fiducial
4P\_fm0.15\_E2\_0.15M\_R14 model. Solid lines correspond to models with
different $M_\mathrm{stripped}$ with $\rho \propto r^{-2}$ while the blue
dashed line corresponds to the model with $\rho \propto
r^{-1}$ and $M_\mathrm{stripped} = 4\times10^{-5}~M_\odot$, as indicated by
the legends in the left panel. In the left panel, the original progenitor
density profile is given by the black dashed line for comparison.
\label{fig:rho}} 
\end{figure*}

\subsubsection{Density profile}\label{sect:rhoprofile}

The CSM density profile is uncertain. If the CSM was created by a
radiation-driven steady wind, it would follow the standard wind density profile
which converges to $r^{-2}$ at sufficiently large $r$.  If the mass ejection
were driven by energy injection from an inner region of the star, the outermost
layers where the binding energy is lowest would be lifted up to make an
outward-moving envelope-like structure. Here we discuss the effect of CSM
density profile on the result. 

As explained above, in our fiducial model we remove a tiny amount of mass
(i.e., $M_\mathrm{stripped}$=$4\times10^{-5}~M_\odot$) from the outermost
layers of the progenitor star and attach the CSM of $\rho(r) \propto r^{-2}$. We
have tested how this choice affects the light curve by adopting different
$M_\mathrm{stripped}$ and $\rho(r)$ as shown in the left panel of
Figure~\ref{fig:rho}.  The total amount of the CSM mass is same for all
different cases (i.e., $M_\mathrm{CSM} = 0.15~M_\odot$).

As shown in the right panel of Figure~\ref{fig:rho}, the difference in
$M_\mathrm{stripped}$ by several orders of magnitude only leads to a
$\sim$0.2 mag difference in the $V$-band light curve during
the IPP. A test model with CSM density profile of $\rho \propto
r^{-1}$ also gives the practically same light curve as the fiducial case of
$\rho \propto r^{-2}$. The differences are negligible compared to the effect of
different CSM mass and radius in our parameter space.
Therefore, the details of a steady-wind-like structure having
$\rho \propto r^{-n}$ with $1 \leq n \leq 2$ play a less important role in the
light curve compared to the CSM mass and radius. It should be kept in mind,
however, that a very different density structure (e.g., a shell-like structure)
might yield a significantly different result during the IPP and that our
calculations would not give a unique solution for the CSM mass and radius.

\subsubsection{CSM radius}\label{sect:Rcsm}

We present  $V$-band light curves and  $B-V$ color evolution of 4P models with
$f_\mathrm{m} = 0.15$ and $E_\mathrm{burst} = 2$B in the upper panels of
Figures~\ref{fig:lig} and~\ref{fig:col}, respectively, for different
combinations of $M_\mathrm{CSM}$ and $R_\mathrm{CSM}$.  It is seen that a
larger CSM radius leads to a brighter emission and a bluer color during the IPP
for a given set of $M_\mathrm{CSM}$, $f_\mathrm{m}$ and $E_\mathrm{burst}$.
For $M_\mathrm{CSM} = 0.15 M_\odot$,  the IPP peak magnitudes of
$M_V =  -16.9, -17.5, -18.4$ are
achieved for $\log R_\mathrm{CSM} \mathrm{[cm]} =$ 13, 14, and 15,
respectively.  This is because a less amount of the kinetic energy is consumed
for the expansion work for a more extended CSM.  

It seems that there exists a certain degree of degeneracy between
$M_\mathrm{CSM}$ and $\log R_\mathrm{CSM}$ with regard to the IPP brightness: a
different set of these parameters could result in a similar IPP peak magnitude.
For example,  in Figure~\ref{fig:lig}, the $V$-band IPP peak of the 4P models
with $M_\mathrm{CSM} = 0.05$ and $\log R_\mathrm{CSM}\mathrm{[cm]} = 14$ looks
fairly comparable to the case with $M_\mathrm{CSM} = 0.15$ and $\log
R_\mathrm{CSM}\mathrm{[cm]} = 13$ (i.e., $M_V \sim -16.8$).  However, the rise
time to the IPP peak becomes longer for a larger $M_\mathrm{CSM}$ mass:
$t_\mathrm{rise} = 3.0$ d for  $M_\mathrm{CSM} = 0.15$ and $\log
R_\mathrm{CSM}\mathrm{[cm]} = 13$ and $t_\mathrm{rise} = 1.9$ d for
$M_\mathrm{CSM} = 0.05$ and $\log R_\mathrm{CSM}\mathrm{[cm]} = 14$.  
Therefore, in principle, the degeneracy between $M_\mathrm{CSM}$ and $\log
R_\mathrm{CSM}$ could be broken, as discussed below in Section~\ref{sect:LSQ14efd}.

\subsubsection{Distribution of \nickel{}}\label{sect:nickel}

The IPP  and the NPP in the light curve can be clearly distinguished if there
exists a sufficient time gap between the two phases.  This time gap becomes
shorter as \nickel{} is more mixed out to the outer layers of the ejecta, which
makes the \nickel{} heating important at an earlier time and the luminosity at
the local minimum of the light curve between the two phases higher (the
bottom-left panel of Figure~\ref{fig:lig}).  For example, in the models of the
figure, the NPP starts  5~d and 2~d  after the IPP peak and the magnitude
difference between the IPP peak and the local minimum is $0.6$ mag and $0.1$ mag
for $f_\mathrm{m}$ = 0.15 and 0.6, respectively.  Therefore, the \nickel{}
mixing is also an important parameter that can significantly interfere the IPP
in the light curve. 

In addition, the overall properties of the NPP are greatly influenced by the
\nickel{} mixing~\citep[e.g., see][for a more detailed discussion]{Yoon2019}.
For example, the NPP peak is reached earlier with a larger \ffm{}.  In the 4P
models presented in Figure~\ref{fig:lig}, it peaks at $t = 18$ d for
$f_\mathrm{m} = 0.6$, and at $t = 22$~d  for $f_\mathrm{m} = 0.15$.  

In the color evolution, the effect of \nickel{} mixing on models with CSM is qualitatively same
with the case of no-CSM, which is discussed in detail by \citet{Yoon2019}. As
mentioned above, a very strong \nickel{} mixing results in a monotonic reddening
(e.g., the case for $f_\mathrm{m} = 0.6$ in the bottom-left panel of
Figure~\ref{fig:col}).  A monotonic reddening is also observed with a
sufficiently large CSM mass (e.g., the case for $M_\mathrm{CSM} = 0.3 M_\odot$
in the top-left panel of Figure~\ref{fig:col}), even when the \nickel{} mixing
is weak (i.e., $f_\mathrm{m} = 0.15$). However, the color during the NPP is
mostly much redder in the former case, for which the \nickel{} abundance and
the resultant opacity at the photosphere are higher.

\subsubsection{Explosion energy}\label{sect:Eburst}

As the explosion energy increases for a given progenitor, the rise time and the
duration of the IPP become shorter (the bottom-right panel of
Figure~\ref{fig:lig}). The time spans between two points at $+ 0.5$ mag from
the IPP peak are 6.4 d, 6.2 d, 6.0 d, respectively for E1.5, E2, E2.5 model
shown in the figure. This is because the expansion velocity of the SN ejecta is
faster, making thermal diffusion more efficient. Also, the luminosity gets
higher during the IPP since a stronger shock creates a hotter and denser
shocked shell.  On the other hand, the higher velocity of the SN ejecta with a
higher explosion energy makes the opacity decrease more quickly, which in turn
makes the recession of the photosphere to the \nickel{} heated reagion faster.
This results in the earlier appearance of the \nickel{} peak for a larger
explosion energy as 27.3 d, 26.2 d, 23.7 d  for the same above models, followed
by a steeper decline as 0.032 mag/d, 0.043 mag/d, 0.047 mag/d during 30 days
from the NPP peak. 

In terms of color, no significant difference during the IPP is found for
different explosion energies in the 4P models (the bottom-right panel of
Figure~\ref{fig:col}). Although the local peak of $B-V$ at the transition
between the IPP and the NPP ($\sim$ 10 d) are similar, the blueward evolution,
which marks the beginning of the NPP, begins somewhat earlier for a higher
explosion energy. Besides color tends to redden more quickly after the NPP peak
because of the faster ejecta cooling.

\subsubsection{Progenitor mass}\label{sect:Mprogenitor}

\begin{figure*}
\includegraphics[width=1\textwidth]{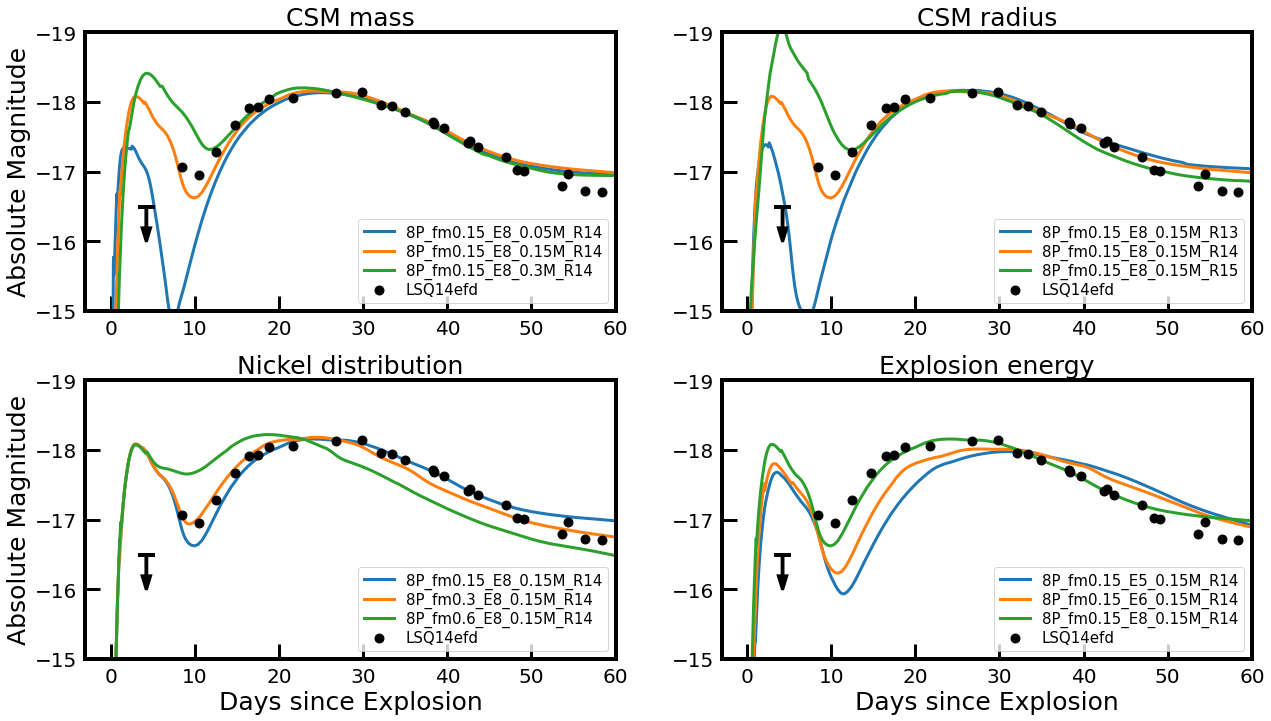}
\caption{Same as in Figure~\ref{fig:lig} but for the 8P SN models. The explosion date is chosen by matching
the observed $V$-band light curve around the main peak with the model 8P\_fm0.15\_E8\_0.15M\_R14.}\label{fig:lig_8P}
\end{figure*}

To explain the light curve width around the NPP peak of the LSQ14efd, we need
$E_\mathrm{burst} \simeq 1.5 - 2.0$B  and $E_\mathrm{burst} \simeq 5.0 - 8.0$B
for the 4P and 8P models~(Figures~\ref{fig:lig} and~\ref{fig:lig_8P}). The
corresponding kinetic energies are $E_\mathrm{K} \simeq 1.2 - 1.7$B and
$E_\mathrm{K} = 3.5 - 6.5$B, respectively.  The higher explosion energies of
the 8P models than the 4P models result in a steeper rise to the IPP peak as
well as a steeper decrease from it.  For our fiducial model, 4P model with
$M_\mathrm{CSM} = 0.15 M_\odot$, $f_\mathrm{m} = 0.15$, $E_\mathrm{burst} = 2$B
and $\log R_\mathrm{CSM} [\mathrm{cm}] = 14$,  the slope of the $V$-band light
curve towards the IPP peak (from +0.5 mag before the peak to the peak) and the
slope beyond the peak (from the peak to the +0.5 mag after the peak) is -0.25
mag/d and 0.11 mag/d. For the corresponding 8P model with $E_\mathrm{burst} =
8$B, the slopes towards the peak and beyond the peak are -0.43 mag/d and 0.14
mag/d, respectively. The IPP peaks in the $V$-band of all 8P models are
brighter by $\sim 0.5$ mag than the 4P models for a given set of the
parameters. 

\begin{figure*}
        \includegraphics[width=1\textwidth]{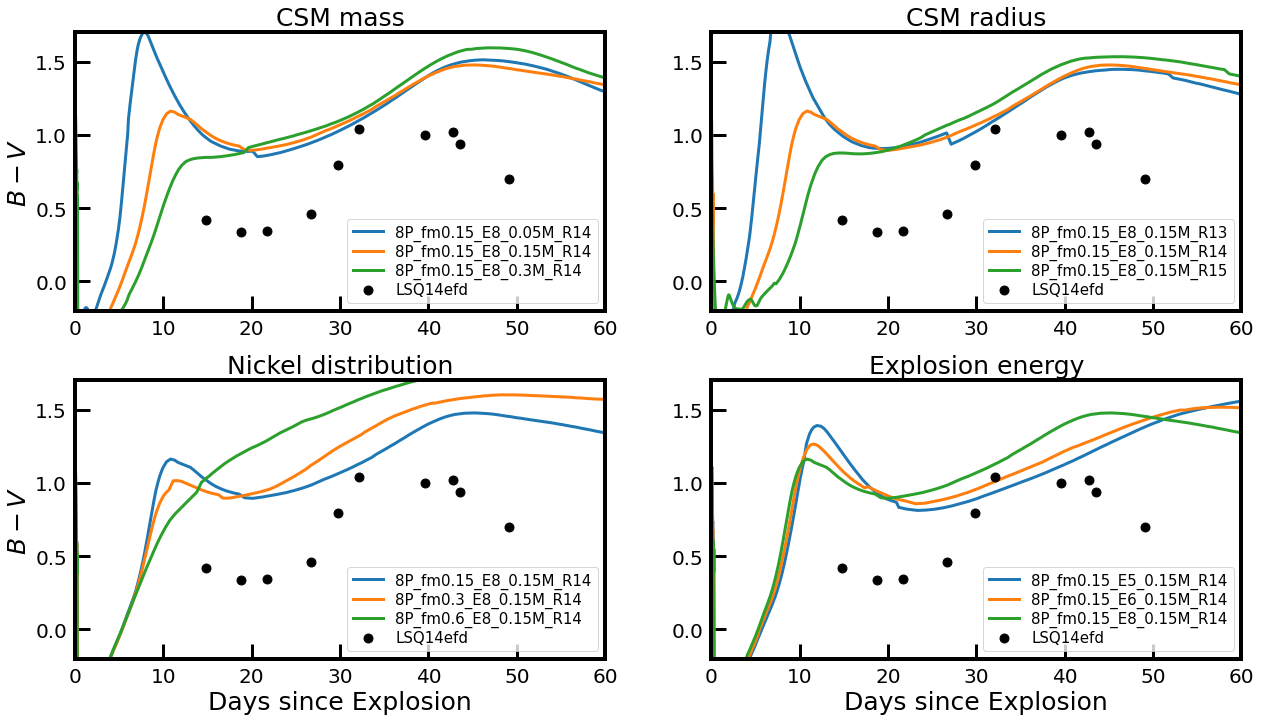}
\caption{Same as in Figure~\ref{fig:col} but for the 8P SN models presented in Figure~\ref{fig:lig_8P}.}
    \label{fig:col_8P}
\end{figure*}

The higher explosion energies make all $B-V$ colors of 8P models redden faster
during the IPP and hence the local peak of $B-V$ at the transition between the
IPP and the NPP larger than in the corresponding cases of 4P models. 
For the 8P models presented in Figure~\ref{fig:col_8P}, 
the local $B-V$ peak at the IPP to NPP transition ($\sim$ 13 d) is 
larger by about 0.7 mag than for the corresponding 4P models (see Figure~\ref{fig:col}). 

On the other hand, the color during the NPP is significantly affected by
\nickel{} heating. The adopted \nickel{} mass is the same for 4P and 8P models
(hence the same amount of \nickel{} heating energy) but the ejecta mass of the
8P models is  2.6 times higher than the 4P models~(see Table~\ref{tab1}).  As a
result, the 8P models are significantly redder during the NPP than the
corresponding 4P models. At the NPP peak, for example, the $B-V$ values of the
8P models are larger by about 0.5 mag than the corresponding 4P models.

\section{Applications to double peaked SNe Ic}

\subsection{LSQ14efd}\label{sect:LSQ14efd}

\begin{figure*}
    \centering
        \includegraphics[width=1.0\textwidth]{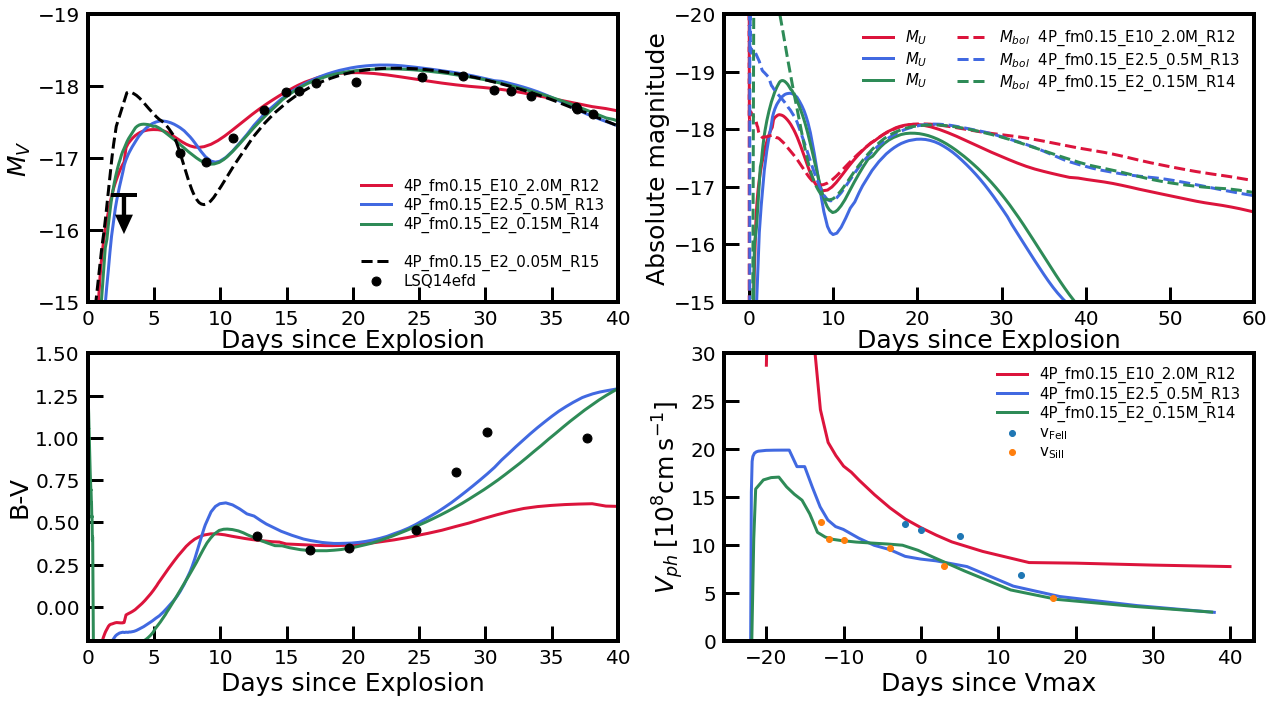}
\caption{\emph{Upper left:} $V$-band light curves of some SN models with different CSM structures (see the labels in the figure) compared to the observed $V$-band light curve of LSQ14efd.  \emph{Lower left}: 
the corresponding $B-V$ color evolution.  \emph{Upper right:} U-band and bolometric light curves of the models.   \emph{Lower right:} Photospheric velocity of the models compared with the FeII and SiII line velocities of LSQ14efd.}
    \label{fig:deg}
\end{figure*}

In this section, we apply our results to the SN Ic LSQ14efd, which motivated
this work. The comparison of our SN models with the observation is done with eyes.   
The observed data are taken from \citet{Barbarino2017}. 
Although a quantitative fitting procedure \citep{Morozova2018,
Ergon2015} is also possible with our grid of models, eye inspection is more than enough
because our grid resolution is rather coarse,  as can be seen in Figures~\ref{fig:lig} and~\ref{fig:lig_8P}

For the comparison of the models with the observation, the
distance modulus of 37.1 is adopted for LSQ14efd.  Given that we have only one
data point of the IPP of this SN, the explosion date cannot be easily
determined from the observation.  
In Figures~\ref{fig:lig} and ~\ref{fig:lig_8P}, 
the explosion date is chosen from the model 
that can best reproduce the $V$-band light  
around the NPP peak. 
The corresponding explosion dates are MJD 56875.5 d and 56874.0~d 
for 4P and 8P models, respectively.  
However, when we make eye inspection to find the best fit model, we freely shift the explosion date 
such that the light curve around the $V$-band NPP peak of each model may match the observation.

As can be seen in Figures~\ref{fig:ubv} and~\ref{fig:lig},
non-detection with the upper magnitude limit of $m_\mathrm{V} = 20.6 \pm 0.2$
at 22 d before the $V$-band maximum is reported for this SN~\citep{Barbarino2017}.
The information about the IPP peak and its rise time is missing and only one
data point before the end of the IPP is available in the $V$-band (i.e., $m_V =
20.03$ at $t = 56882.48$ MJD; the first observed point).  The local minimum in the $V$-band at the
trasition between IPP and NPP ($t = 56884.44$ MJD; the second observed point) is $m_V = 20.15$. The data
in the other optical bands are available only after the IPP. 
 
To find the model parameters that can give a consistent fit to LSQ14efd, we
take the following steps. First, the amount of \nickel{} is fixed to
$0.25~M_\odot$, which is inferred from the NPP peak
brightness~\citep{Barbarino2017}, as explained above. Second, we only use the
models with $f_\mathrm{m} = 0.15$ because the color evolution of these models
is qualitatively same as that of LSQ14efd where the signature of relatively
weak \nickel{} mixing is found (see the discussion in
Section~\ref{sect:nickel}).

For the remaining sets, we search for the CSM parameters (i.e., $\log
R_\mathrm{CSM}$ and $M_\mathrm{CSM}$) that can best explain the three data
points of the observed IPP (i.e, the non-detection limit, the first observed
point, and the second observe point which is  the local minimum at the IPP to
NPP transition; see Figure~\ref{fig:lig}) as well as $E_\mathrm{burst}$ that
can give a reasonable $V$-band light curve width compared to LSQ14efd.  

We find that no 8P model can satisfy the non-detection limit: given the very
high energies (i.e,. $E_\mathrm{burst} = $ 6.0B and 8.0B), too bright emission
is predicted at the point of non-detection for the models that can match the
second and third data points of the IPP (see Figure~\ref{fig:lig_8P}). Note
also that the color predicted by the 8P models is much redder than the
observation (i.e., by more than 0.6 mag at the NPP peak in terms of $B-V$;
Figure~\ref{fig:col_8P}).  

Within the grid of our fiducial 4P models  with $f_\mathrm{m} = 0.15$, we find
that the IPP brightness can be best reproduced by the model with
$M_\mathrm{CSM}  = 0.15 M_\odot$, $R_\mathrm{CSM} = 10^{14}~\mathrm{cm}$ and
$E_\mathrm{burst} = 2.0$B. This best fit model is presented in
Figure~\ref{fig:deg} (the green line).  The model with $R_\mathrm{CSM} =
10^{15}$~cm which can well reproduce the first observed point is also shown in
the figure for comparison (the dashed line). This model predicts too faint
emission at the second point of the observed IPP, and can be ruled out. 

As discussed in Section~\ref{sect:Rcsm}, different combinations of
$M_\mathrm{CSM}$ and $R_\mathrm{CSM}$ can yield a similar IPP peak.  To
investigate this uncertainty, in Figure~\ref{fig:deg}, we  also present two
test models for which we adopt $M_\mathrm{CSM} = 2.0 M_\odot$  and
$R_\mathrm{CSM} = 10^{12}$~cm  (the red line) and $M_\mathrm{CSM} = 0.5
M_\odot$ and $R_\mathrm{CSM} = 10^{13}$~cm (the blue line).  These models and 
our best fit model have
a similar IPP peak in $V$-band. However, the evolution
after the IPP peak is different for each case.  The model with $M_\mathrm{CSM}
= 2.0 M_\odot$ and $R_\mathrm{CSM} = 10^{12}$~cm have a very long term effect
of CSM and predicts brighter emission at the second and third observed points.
The $B-V$ color evolution of this model after the NPP peak is also
distinctively different from the observation.  The model with  $M_\mathrm{CSM}
= 0.5 M_\odot$ and $R_\mathrm{CSM} = 10^{13}$~cm has a longer decline rate from
the IPP peak than our best fit model and predict too bright emission compared
to the observation at the first observed point. 

We find that the degeneracy between  $M_\mathrm{CSM}$ and $R_\mathrm{CSM}$
could be more easily broken in the U band as seen in the upper right panel of
Figure~\ref{fig:deg}.   The compared tree models have a similar IPP peak in the
V band but the U band IPP peak is systematically brighter for a larger
$R_\mathrm{CSM}$, for which the adiabatic cooling is less efficient.
Therefore, U-band observations during early times would be most useful in
future observational studies on the CSM properties. 

In principle, the degeneracy between $M_\mathrm{ej}$ and
$E_\mathrm{K}$ which exists when inferring SN parameters with a NPP light curve
can be broken by comparing the photospheric velocity and the model prediction.
As seen in Figure~\ref{fig:deg}, the photospheric velocity evolution of our
best fit model is consistent with  the observed $v_\mathrm{Si II}$.
\citet{Barbarino2017} use $v_\mathrm{Fe II}$, which is higher than $v_\mathrm{Si
II}$, to infer SN parameters and as a result obtain higher $M_\mathrm{ej}$ 
and $E_\mathrm{K}$ (i.e., $6.3~\mathrm{M_\odot}$ and 5.6B) than our fiducial values. 
However, as discussed above, such a high SN energy of 5.6B is not favored when 
the observed IPP is compared with the models. 
In our models, the photosphere is defined by the Rosseland mean opacity 
and it would be an interesting subject
of future work to investigate which absorption line better traces the
Rosseland mean photosphere.

We conclude that the early-time light curve of LSQ14efd is consistent with the
SN model prediction with  a massive CSM of about $M_\mathrm{CSM} \approx 0.15
M_\odot$ extending up to about $R_\mathrm{CSM} \approx 10^{14}$~cm.  This
corresponds to a mass loss rate of $\dot{M} \approx 1.0
M_\odot~\mathrm{yr^{-1}}$ during $\sim 0.2$~yr before the SN explosion if we
assume the terminal wind velocity as $200~\mathrm{km~s^{-1}}$. We discuss its
implications for the mass loss mechanism below in Section~\ref{sect:mechanism}.

\subsection{iPTF15dtg and SN 2020bvc}\label{sect:dtg,bvc}

\begin{figure*}
    \centering
    \includegraphics[width=1\textwidth]{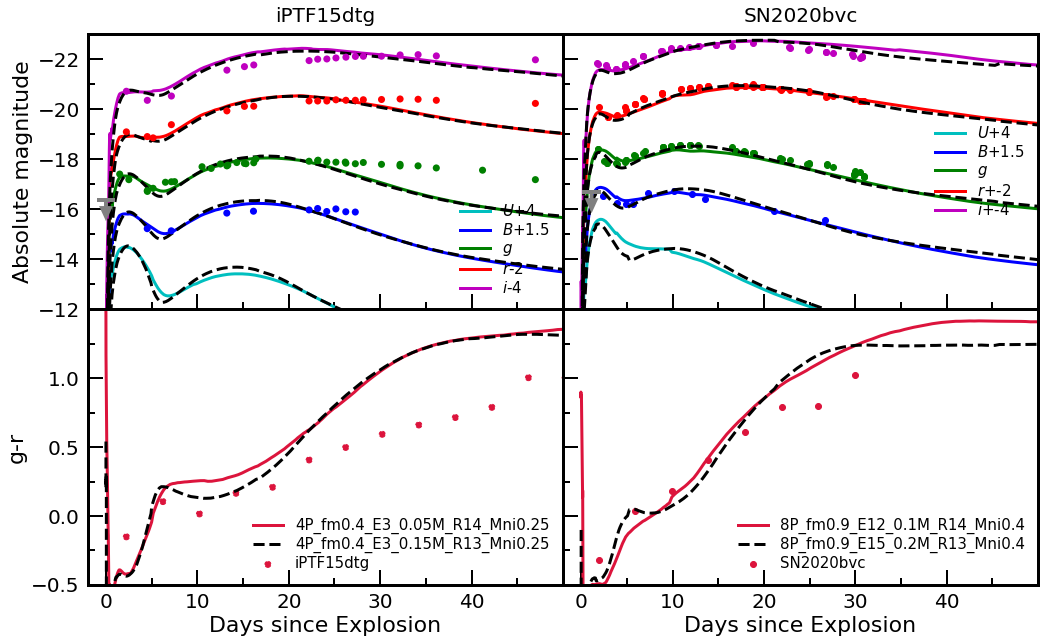}
\caption{Light curves and color evolution of iPTF15dtg and SN2020bvc overlapped with the best-fit models. In the upper panels, solid lines and dashed lines represent R14 models and R13 models, respectively. Bottom panels show $g-r$ color evolution of the same models. $g-r$ of each SN is obtained from linear interpolation of each band magnitude. Photometric data and extinction correction of iPTF15dtg were imported from The Open Supernova Catalog and \citet{Taddia2016}, and SN2020bvc from \citet{Ho2020}. For iPTF15dtg, the explosion date is chosen to be MJD 57332.9 d, the last non-detection date. For SN 2020bvc, it is chosen by matching the observed $V$-band light curve around the main peak with our best fit model. Grey downward arrows indicate the pre-explosion limit in g-band.}\label{fig:ptf,bvc}
\end{figure*}

We also compare our models with two other double peaked SNe Ic SN iPTF15dtg and
SN 2020bvc.  iPTF15dtg is a peculiar SN Ic which is suspected to be powered by a
magnetar~\citep{Taddia2019} and its optical light curves around the main peak
cannot be easily explained by our grid of models. Given that
our main interest is to infer the properties of CSM, here we do not attempt to make
a model that can reproduce the NPP \citep[see instead][who inferred SN
parameters from the light curve around the main peak]{Taddia2016}. As discussed above, the 
IPP light curve is largely determined by $M_\mathrm{CSM}$, $R_\mathrm{CSM}$ and $E_\mathrm{burst}$, 
which can be fairly well determined independently of the detailed properties of the NPP 
if the early time data of the IPP are good enough.

Comparison of our model grid with the IPP light curve of this SN indicates that
the early-time light curve (0 $\sim$ 15 d
after the explosion) of iPTF15dtg is fairly consistent with two of our 4P
models: 4P\_fm0.4\_E3\_0.05M\_R14 and
4P\_fm0.4\_E3\_0.15M\_R13, as seen in Figure~\ref{fig:ptf,bvc}.
Here, the last
non-detection date is chosen for the explosion date since we cannot fit the main peak with our grid of models.
This implies that the progenitor of iPTF15dtg had CSM with $R_\mathrm{CSM}=
10^{13} -  10^{14}\mathrm{cm}$ and $M_\mathrm{CSM}= 0.05 - 0.15 M_\odot$.

For SN 2020bvc, we already presented  a result of our model comparison with the
observation in another paper ~\citep[][]{Rho2020}.  For this particular SN, we
use the 8P progenitor and assume that $^{56}$Ni is uniformly distributed in the
inner 90\% of the SN ejecta.  The \nickel{} mass required to explain the NPP
peak  is found to be $0.4~M_\odot$. As in the case of LSQ14efd,
the explosion date is determined by matching the observed $V$-band light curve around
the main peak with the models.    For the
properties of these models made for the comparison with SN 2020bvc, see
Table~\ref{tab3}.  Within our grid, we find that the overall light curve
properties including the IPP of SN 2020bvc are most consistent with the
following two sets of parameters: $M_\mathrm{CSM} = 0.1 M_\odot$,
$R_\mathrm{CSM} = 10^{14}~\mathrm{cm}$ \&  $E_\mathrm{burst} = 12$~B, and
$M_\mathrm{CSM} = 0.2 M_\odot$, $R_\mathrm{CSM} = 10^{13}~\mathrm{cm}$ \&
$E_\mathrm{burst} = 15$~B.  The early-time color of this SN is very blue as
predicted by the model, which is due to the SN interaction with CSM.

\subsection{Implications for the CSM formation mechanism}\label{sect:mechanism}

Our result indicates that the inferred CSM properties of the double peaked SNe
Ic considered in this study  (LSQ14efd, iPTF15dtg and SN 2020bvc) are
intriguingly similar:  $M_\mathrm{CSM} \approx 0.1-0.2 M_\odot$ and
$R_\mathrm{CSM} \approx 10^{13} - 10^{14}~\mathrm{cm}$.

The implied mass loss rate is $\dot{M} \approx 0.6 - 13.0$ $(v_\mathrm{w}/200~
\mathrm{km~s^{-1}}) ~M_\odot~\mathrm{yr^{-1}}$. The mass eruption should have
occurred within about 0.2 yr from the explosion if we adopt $v_\mathrm{w} =
200~\mathrm{km~s^{-1}}$.  

This high mass loss rate inferred in our study cannot be explained by the
conventional line-driven wind of Wolf-Rayet stars.  \citet{Fuller2018} explore
the possibility of pre-supernova outbursts via wave heating  during core neon
and oxygen burning in a 5~$M_\odot$ hydrogen-free helium star.  The predicted
mass loss rate is \wave0.01$M_\odot~\mathrm{yr^{-1}}$ with a wind terminal
velocity of a $450~\mathrm{km~s^{-1}}$. This is smaller by  two or three orders of
magnitude than our inferred value.  Note also that helium-poor SN Ic progenitor
stars are supposed to be more compact by several factors than helium-rich SN Ib
progenitors which are considered by Fuller et al. ~\citep[e.g.,][]{Yoon2019}.
Therefore, it seems that the inferred CSM property of the double peaked SNe Ic
cannot be easily explained by the wave heating model. 

\citet{Aguilera2018} find that the surface layers of helium stars could be
spun-up to the critical rotation during the final evolutionary stages if the
helium stars retained sufficiently high angular momenta.  The predicted mass
loss rate is about $ \dot{M} \lesssim 0.01 M_\odot~\mathrm{yr^{-1}}$ during the
last years of the evolution, which is also much lower than our
inferred mass loss rate. 

\citet{Woosley2019} find that very massive CSM of $0.02 \cdots 0.74~M_\odot$
can be created by silicon flash for helium stars having initial helium star
masses of 2.5 - 3.2~$M_\odot$. However, all these He star models with silicon
flash have a fairly  massive helium envelope ($ > 0.7 M_\odot$) and the
resulting SN would be a SN Ib rather than SN Ic.  In addition, the silicon
flash only occurs for a relatively small progenitor mass (i.e,
$M_\mathrm{final} \lesssim 2.6 M_\odot$), which could not easily explain 
the light curves of the double peaked SNe Ic of our sample.  

One scenario that could explain the CSM of the double peaked SNe Ic of our
sample would be the possibility that the mass loss from the progenitor was
induced by the combined effects of wave heating and rotation.  If the
progenitor were rotating at the critical rotation during the core neon and
oxygen burning stages, the binding energy of the outermost layers would be
lower than the non-rotating case and the mass loss due to the wave heating
would be easier.  

This scenario is in line with the fact that iPTF15dtg and SN 2020bvc belong to
peculiar SNe Ic (ie., magnetar-powered/broad-lined SNe Ic), for which a rapid
rotating progenitor is often invoked. This  might also imply that the LSQ14efd
progenitor was a rapid rotator even though the inferred properties of LSQ14efd
do not look peculiar except for the IPP signature.  It is possible
that LSQ14efd was in part powered by a mangetar, in which case the actual
\nickel{} would be smaller than the inferred value of $0.25~M_\odot$.  

Another possibility is that the progenitors did not really undergo a mass
eruption, but simply an expansion of the outer layers due to an energy
injection during the pre-SN stage.  Althernatively, the IPP observed in our
sample might not be related to CSM, but to interactions with a companion
star~\citep{Kasen2010} or to high velocity \nickel{} due to an asymmetric
explosion~\citep[e.g.,][]{Folatelli2006, Bersten2013}.   An elaboration of
these scenarios would be a subject of future work.

\section{Conclusions}

We have discussed the effects of CSM on the early-time SNIc light curve and color evolution.  SN models with different CSM mass, CSM radius, fm, and Eburst are investigated in a systematic fashion to understand the IPP properties.  In Table~\ref{tab2}, we present the IPP properties of the investigated SN models for \ffm=0.15, 
which can be summarized as follows. 
\begin{enumerate}
\item Models with more massive CSM have brighter IPP peaks in the optical bands for a given initial condition. For example, for all 4P models given in Table~\ref{tab2}, the $V$-band peak during the IPP gets higher by -0.32mag when CSM mass increases by 0.1$M_\odot$ (values obtained by linear regression) due to efficient conversion of kinetic energy into thermal energy (Section~\ref{sect:Mcsm}). At the same time, the rising time ($t_\mathrm{rise}$) and the light curve width ($\Delta t_{+0.3}$) get extended by 1.61 d and 1.19 d due to a longer diffusion time scale. 
\item The CSM radius has the same qualitative effect on IPP properties as the CSM mass. Models with a larger CSM radius consume less energy due to the expansion work thus making the IPP brighter (Section~\ref{sect:Rcsm}). For example, 4P model in Table~\ref{tab2} gets brighter by -0.13~mag when its CSM radius increases by $10^{14}$~cm. It also makes the rising time and the width of the IPP longer, by 0.20~d and 0.16~d.
\item  A higher explosion energy makes the IPP brighter and its time scale shorter (Section~\ref{sect:Eburst}). For our 4P models, the IPP peaks higher by -0.63~mag, the rising time and the IPP duration are shortened by -1.28~d and -0.96~d when their explosion energy is increased by 1B, respectively.  However, it has a negligible effect on the color evolution during the IPP for a given explosion condition within our considered parameter space. 
\item The IPP can be significantly interfered by  $^{56}$Ni heating if $^{56}$Ni mixing is sufficiently strong (Section~\ref{sect:nickel}). In particular, the local brightness minimum between IPP and NPP becomes larger and the time span between the IPP and NPP peaks shorter with a stronger $^{56}$Ni mixing.  
\item The early-time color becomes significantly bluer with CSM compared to the case without CSM. 
\end{enumerate}

We compare our models with three double-peaked SNe Ic LSQ14efd, iPTF15dtg and
SN 2020bvc. We find that the inferred CSM properties of these SNe Ic are
intriguingly similar:  a CSM mass of $M_\mathrm{CSM}  = \sim 0.1  M_\odot$ and
a CSM radius of $R_\mathrm{CSM}  = 10^{13} - 10^{14}~\mathrm{cm}$.  This
possibly suggests a common  mechanism for the CSM formation from the progenitors of these
SNe Ic.  The implied mass loss rate of $\dot{M} \gtrsim
1.0~M_\odot~\mathrm{yr^{-1}}~(v_\mathrm{w}/200~\mathrm{km/s})$ seems to be too
high to be explained by the existing theories such as wave heating or
rotationally-induced mass shedding (Section~\ref{sect:mechanism}).  Future work
needs to address if there exists a possible mechanism to explain such massive
extended material around the progenitor star, or if an alternative scenario
such as asymmetric explosion might explain the bright IPP of the double peaked
SNe Ic of our sample.

\acknowledgments

This work has been supported by the National Research Foundation of Korea (NRF)
grant (NRF-2019R1A2C2010885).  We are grateful to Taebum Kim for creating the
python package for the Kippenhahn diagram and to Wonseok Chun for providing the
progenitor models to us. Work by S.B. on the development of STELLA code is
supported by the Russian Science Foundation grant 19-12-00229.

\begin{table*}[]
\caption{IPP properties of the models for LSQ14efd with \ffm = 0.15}\label{tab2}
\begin{center}
\begin{tabular}{ccccccc|ccccccc}
\hline
\multicolumn{7}{c|}{4P progenitor}                                                                                                                    & \multicolumn{7}{c}{8P progenitor}                                                                                                                     \\
$E_\textrm{burst}$ & $M_\textrm{CSM}$ & log$R_\textrm{CSM}$ & $M_{V, \mathrm{peak}}$ & $t_\mathrm{rise}$ & $\Delta t_{+0.3}$ & s$_\mathrm{3\sim 10d}$ & $E_\textrm{burst}$ & $M_\textrm{CSM}$ & log$R_\textrm{CSM}$ & $M_{V, \mathrm{peak}}$ & $t_\mathrm{rise}$ & $\Delta t_{+0.3}$ & s$_\mathrm{3\sim 10d}$ \\
{[}B{]} & {[}$M_\odot${]}  & {[}cm{]}            & {[}mag{]}              & {[}d{]}           & {[}d{]}           & {[}mag/d{]}            & {[}B{]} & {[}$M_\odot${]}  & {[}cm{]}            & {[}mag{]}              & {[}d{]}           & {[}d{]}           & {[}mag/d{]}            \\ \hline \hline
1                  & 0.05             & 13                  & -15.79                 & 2.35              & 1.75              & 0.29                   & 5                  & 0.05             & 13                  & -16.53                 & 1.42              & 1.35              & 0.38                   \\
                   &                  & 14                  & -16.26                 & 2.89              & 2.76              & 0.21                   &                    &                  & 14                  & -17.06                 & 1.67              & 2.89              & 0.40                   \\
                   &                  & 15                  & -17.23                 & 4.05              & 3.46              & 0.08                   &                    &                  & 15                  & -18.26                 & 2.82              & 1.85              & 0.31                   \\ \cline{2-7} \cline{9-14} 
                   & 0.15             & 13                  & -16.26                 & 3.88              & 2.84              & 0.32                   &                    & 0.15             & 13                  & -17.10                 & 2.66              & 2.06              & 0.45                   \\
                   &                  & 14                  & -16.79                 & 5.39              & 4.79              & 0.08                   &                    &                  & 14                  & -17.68                 & 3.57              & 3.26              & 0.17                   \\
                   &                  & 15                  & -17.82                 & 6.71              & 6.18              & 0.04                   &                    &                  & 15                  & -18.73                 & 4.78              & 2.38              & 0.07                   \\ \cline{2-7} \cline{9-14} 
                   & 0.3              & 13                  & -16.55                 & 5.71              & 3.66              & 0.16                   &                    & 0.3              & 13                  & -17.33                 & 3.81              & 2.89              & 0.32                   \\
                   &                  & 14                  & -17.16                 & 7.93              & 6.36              & 0.04                   &                    &                  & 14                  & -18.04                 & 5.17              & 3.84              & 0.09                   \\
                   &                  & 15                  & -18.09                 & 9.19              & 9.15              & 0.01                   &                    &                  & 15                  & -18.92                 & 6.24              & 3.68              & 0.02                   \\ \hline
1.5                & 0.05             & 13                  & -16.17                 & 1.79              & 1.55              & 0.20                   & 6                  & 0.05             & 13                  & -16.68                 & 1.34              & 1.36              & 0.31                   \\
                   &                  & 14                  & -16.59                 & 2.13              & 3.20              & 0.21                   &                    &                  & 14                  & -17.23                 & 1.35              & 2.83              & 0.38                   \\
                   &                  & 15                  & -17.70                 & 3.59              & 2.33              & 0.14                   &                    &                  & 15                  & -18.48                 & 2.88              & 1.67              & 0.30                   \\ \cline{2-7} \cline{9-14} 
                   & 0.15             & 13                  & -16.66                 & 3.42              & 2.67              & 0.28                   &                    & 0.15             & 13                  & -17.18                 & 2.46              & 2.09              & 0.38                   \\
                   &                  & 14                  & -17.18                 & 4.92              & 4.63              & 0.10                   &                    &                  & 14                  & -17.80                 & 3.31              & 3.55              & 0.19                   \\
                   &                  & 15                  & -18.19                 & 5.85              & 4.08              & 0.04                   &                    &                  & 15                  & -18.92                 & 4.57              & 2.19              & 0.07                   \\ \cline{2-7} \cline{9-14} 
                   & 0.30             & 13                  & -16.95                 & 4.92              & 3.40              & 0.19                   &                    & 0.30             & 13                  & -17.49                 & 3.61              & 2.69              & 0.32                   \\
                   &                  & 14                  & -17.57                 & 6.68              & 5.40              & 0.05                   &                    &                  & 14                  & -18.18                 & 4.93              & 3.84              & 0.09                   \\
                   &                  & 15                  & -18.42                 & 7.87              & 6.47              & 0.02                   &                    &                  & 15                  & -19.07                 & 6.00              & 3.10              & 0.02                   \\ \hline
2                  & 0.05             & 13                  & -16.39                 & 1.61              & 1.49              & 0.15                   & 8                  & 0.05             & 13                  & -16.88                 & 1.20              & 1.23              & 0.22                   \\
                   &                  & 14                  & -16.83                 & 1.88              & 3.18              & 0.21                   &                    &                  & 14                  & -17.37                 & 2.54              & 2.94              & 0.35                   \\
                   &                  & 15                  & -17.92                 & 3.14              & 2.49              & 0.16                   &                    &                  & 15                  & -18.66                 & 2.75              & 1.57              & 0.28                   \\ \cline{2-7} \cline{9-14} 
                   & 0.15             & 13                  & -16.92                 & 2.97              & 2.53              & 0.26                   &                    & 0.15             & 13                  & -17.41                 & 2.64              & 2.03              & 0.38                   \\
                   &                  & 14                  & -17.45                 & 4.14              & 4.29              & 0.11                   &                    &                  & 14                  & -18.08                 & 2.89              & 2.97              & 0.20                   \\
                   &                  & 15                  & -18.42                 & 5.40              & 3.51              & 0.05                   &                    &                  & 15                  & -19.10                 & 4.13              & 1.76              & 0.07                   \\ \cline{2-7} \cline{9-14} 
                   & 0.3              & 13                  & -17.16                 & 4.44              & 3.52              & 0.19                   &                    & 0.3              & 13                  & -17.75                 & 3.16              & 2.56              & 0.31                   \\
                   &                  & 14                  & -17.87                 & 6.16              & 4.91              & 0.06                   &                    &                  & 14                  & -18.41                 & 4.24              & 3.67              & 0.11                   \\
                   &                  & 15                  & -18.62                 & 7.32              & 6.35              & 0.02                   &                    &                  & 15                  & -19.26                 & 5.47              & 3.47              & 0.03                   \\ \hline
\end{tabular}
\tablecomments{$E_\textrm{burst}$: explosion energy; $M_\textrm{CSM}$: mass of the CSM; log$R_\textrm{CSM}$: radius of the CSM; $M_{V, \mathrm{peak}}$: $V$-band maximum magnitude during the IPP; $t_\mathrm{rise}$: time from the explosion to the $V$-band maximum during the IPP, $\Delta t_{+0.3}$: time span between two 76\% V-mag maximum (+0.3 mag) points during the IPP, one during its rise to the maximum and the other during its decline from the maximum; s$_\mathrm{3\sim 10d}$: $B-V$ color evolution slope at 3\wave10 day.} 
\end{center}
\end{table*}

\begin{table*}[]
\caption{IPP properties of the models for SN2020bvc with \ffm = 0.9}\label{tab3}
\begin{tabular}{lllllll}
\hline
$E_\textrm{burst}$ & $M_\textrm{CSM}$ & log$R_\textrm{CSM}$ & $M_{g, \mathrm{IPP\,peak}}$ & $t_\mathrm{rise}$ & $\Delta t_{+0.3}$ & s$_\mathrm{3\sim 10d}$ \\
{[}B{]} & {[}$M_\odot${]}  & {[}cm{]}            & {[}mag{]}              & {[}d{]}           & {[}d{]}           & {[}mag/d{]}            \\ \hline
6       & 0.05 & 13 & -16.91 & 1.37 & 1.20 & 0.01 \\
        &      & 14 & -17.38 & 1.79 & 2.48 & 0.06 \\
        &      & 15 & -18.52 & 2.81 & 2.23 & 0.05 \\ \cline{2-7} 
        & 0.1  & 13 & -17.20 & 1.94 & 1.67 & 0.04 \\
        &      & 14 & -17.88 & 2.52 & 2.28 & 0.07 \\
        &      & 15 & -18.88 & 3.84 & 1.65 & 0.06 \\ \cline{2-7} 
        & 0.2  & 13 & -17.49 & 2.75 & 2.40 & 0.06 \\
        &      & 14 & -18.23 & 3.85 & 3.05 & 0.06 \\
        &      & 15 & -19.14 & 5.06 & 2.34 & 0.04 \\ \hline
9       & 0.05 & 13 & -17.21 & 1.14 & 1.09 & 0.00  \\
        &      & 14 & -17.66 & 1.74 & 2.31 & 0.06 \\
        &      & 15 & -18.83 & 2.75 & 2.08 & 0.07 \\ \cline{2-7} 
        & 0.1  & 13 & -17.54 & 2.13 & 1.57 & 0.04 \\
        &      & 14 & -18.19 & 2.32 & 2.42 & 0.08 \\
        &      & 15 & -19.29 & 3.41 & 1.30 & 0.07 \\ \cline{2-7} 
        & 0.2  & 13 & -17.88 & 2.41 & 1.9  & 0.06 \\
        &      & 14 & -18.57 & 3.24 & 2.85 & 0.07 \\
        &      & 15 & -19.43 & 4.42 & 1.21 & 0.04 \\ \hline
12      & 0.05 & 13 & -17.39 & 1.05 & 1.04 & 0.02 \\
        &      & 14 & -17.92 & 2.14 & 2.02 & 0.07 \\
        &      & 15 & -19.02 & 2.50 & 2.20 & 0.09 \\ \cline{2-7} 
        & 0.1  & 13 & -17.75 & 1.56 & 1.43 & 0.04 \\
        &      & 14 & -18.38 & 2.17 & 2.19 & 0.08 \\
        &      & 15 & -19.50 & 3.24 & 1.15 & 0.07 \\ \cline{2-7} 
        & 0.2  & 13 & -18.10 & 2.48 & 1.88 & 0.07 \\
        &      & 14 & -18.81 & 3.01 & 2.64 & 0.07 \\
        &      & 15 & -19.65 & 3.91 & 1.35 & 0.05 \\ \hline
\end{tabular}
\tablecomments{Column names are same as Table~\ref{tab2} except for $M_{g, \mathrm{peak}}$: $g$-band maximum magnitude during the IPP; $t_\mathrm{rise}$: time from the explosion to the $g$-band maximum during the IPP, $\Delta t_{+0.3}$: time span between two 76\% g-mag maximum (+0.3 mag) points during the IPP, one during its rise to the maximum and the other during its decline from the maximum; s$_\mathrm{3\sim 10d}$: $g-r$ color evolution slope at 3\wave10 day.}
\end{table*}

\bibliography{main}{}
\end{document}